\documentstyle[12pt,aps]{revtex}
\begin{document}
\draft

\def\beginwide{
        \end{multicols} \vspace*{-0.5cm} \noindent
        \rule{3.5in}{.1mm}\rule{.1mm}{5mm} \widetext \medskip }
\def\beginwidetop{
        \end{multicols} \vspace*{-0.5cm} \noindent
        \widetext \medskip }
\def\endwide{
        \hspace*{3.35in}~\rule[-5mm]{.1mm}{5mm}\rule{3.5in}{.1mm}
        \begin{multicols}{2} \vspace*{-1.0cm} \noindent }
\def\endwidebottom{
        \begin{multicols}{2} \vspace*{-1.0cm} \noindent }

\newcommand{\beq}{\begin{equation}}
\newcommand{\eeq}{\end{equation}}
\newcommand{\bdis}{\begin{displaymath}}
\newcommand{\edis}{\end{displaymath}}
\newcommand{\bea}{\begin{eqnarray}}
\newcommand{\eea}{\end{eqnarray}}
\newcommand{\barr}{\begin{array}}
\newcommand{\earr}{\end{array}}
\title{Exact velocity of dispersive flow in the asymmetric
avalanche process}
\author{E.V. Ivashkevich, A.M. Povolotsky, V.B. Priezzhev}
\address{
Bogoliubov Laboratory of Theoretical Physics,
Joint Institute for Nuclear Research, Dubna 141980, Russia\\ }
\date{\today}
\maketitle
\begin{abstract}
Using the Bethe ansatz we obtain the exact solution for the
one-dimensional asymmetric avalanche process. We evaluate the
velocity of dispersive flow as a function of driving force and the
density of particles. The obtained solution shows a dynamical
transition from  intermittent to continuous flow.
\end{abstract}
\vspace{0.3cm}
\hspace{1.5cm}

\section{Introduction and the model definition.}

The avalanche dynamics is a basic scenario of relaxation of
unstable states in extremal systems  where each movable element is
near a border of stability. A typical long-tailed distribution of
avalanche sizes leads to the dispersive transport of particles
\cite{Today}. As an illustrative example, granular systems exhibit
intermittent avalanches which enables one to use granular piles
(sand piles, rice piles) for explanation of self-organized
criticality in generic dissipative systems \cite{BTW}. In the past
decade, it has become clear that the dispersive transport  can be
recast in terms of interface depinning \cite{KS,TKarDhar} and
various growth models \cite{PacBoet}. Recently, a dynamical
transition from intermittent to continuous flow in a random
sandpile model has been revealed \cite{CorPac}. Nevertheless, an
explicit theoretical description of stochastic avalanche processes
and an exact evaluation of characteristics of dispersive flow
remains an open problem.

Despite drastic simplifications which were introduced to mimic
real avalanches, exact results are scarce even for the
determimistic dynamics. As to stochastic dynamics, it is
especially difficult as it is beyond the class of abelian models
\cite{D}, where asymmetric processes  appear to be solvable
\cite{DharRam}. The situation is to be compared with the theory of
exclusion processes where many properties, such as steady states,
average current, diffusion constant etc. have been calculated for
an asymmetric one-dimensional case \cite{GwaSp,DerLeb,Derrida,KK}.
The usual presentation of the asymmetric exclusion process (ASEP)
is given by a master equation for the probability
$P_t(x_1,...,x_P)$ of finding $P$ particles at time $t$ on sites
$x_1,...,x_P$ of a ring consisting of $N$ sites. During any time
interval $dt$, each particle jumps with probability $dt$ to its
right if the target site is empty. This elementary restriction
leads to a non-trivial problem of evaluation of the steady state
properties, which can be solved by the Bethe ansatz.

In a similar way, the simplest asymmetric avalanche process (ASAP)
can be formulated as follows. In a stable state, each of $N$ sites
on a ring is either occupied by one particle or empty. The total
number of particles $P$ is fixed.  During time interval $dt$, each
particle jumps with probability $dt$ to its right. In the course
of time, some site $x$ may get unstable with occupation number
$n>1$. Then it must relax immediately to the stable state by
transferring to its right either $n$ particles with probability
$\mu_n$ or $n-1$ particles with the probability $1-\mu_n$. The
quantity $\mu_n$ can be associated with a driving force acting on
the unstable group of $n$ particles.

The main difference between the ASEP and ASAP lies in the depth of
reconstruction of a configuration $C=\{x_1,...,x_P\}$ during the time
interval $dt$. In the ASEP, the total distance $Y_t$ covered by all
particles between time $0$ and $t$ increases by $1$ during $dt$ if the
configuration $C$ differs from a new one $C^{'}$ or remains unchanged
if the motion is forbidden. In the ASAP, the motion of a particle is
always possible and increase of $Y_t$ is not bounded. Thus, the
configuration $C$ may be completely different from $C^{'}$ depending on
numbers of particles spilled to right from each unstable site.

The present formulation of the ASAP is inspired by works \cite{Vesp}
where a model of activated random walks is introduced and  \cite{MZhang}
where the directed avalanche dynamics is formulated in terms of continuous
variables. Under an assumption about independence of variations of the
avalanche size at each time step, the probability distribution of avalanche
sizes is found exactly \cite{MZhang},\cite{F}. However, the configurational
space of the continuous model is too complicated to determine steady
state features.

Here, extending the Bethe anzatz approach to exclusion processes, we
obtain the expression for the generating function of $Y_t$ for the discrete
ASAP in the thermodynamic limit of large $N$ for a fixed density of
particles $\rho=P/N$. We find two phases corresponding to a dispersive
flow  and a continuous flow, and evaluate the exact average velocity
in the whole range of parameters of the first phase. We determine the
separation line between two phases where avalanches are critical.

\section{The dynamical rules and the average velocity.}
Even before going into details of our calculations, we can get
important restriction on the toppling rules of the ASAP. For the
problem to be solvable by Bethe ansatz, one has to make sure that
many-particle problem can be reduced to the problem of two
interacting particles. In other words, the toppling probabilities
$\mu_n$ of unstable configuration with $n>1$ particles at the same
site should be determined recursively in terms of two-particle
toppling probability $\mu_2$ only. Namely, the probability for $n$
particle to leave the site is the sum of two processes. In the
first process two particles leave the site with probability
$\mu_2$ and then $n-2$ particles leave the site with probability
$\mu_{n-2}$. The second process corresponds to spilling two
particles with probability $1-\mu_2$ when only one particle leave
the site and then remaining $n-1$ particles leave the site with
probability $\mu_{n-1}$. Thus, we obtain the recursion relations
which express all the probabilities trough the only constant
$\mu$
\begin{eqnarray}
\mu_1&=&0\nonumber\\ \mu_2&=&\mu,\nonumber\\
\mu_n&=&\mu_{n-2}\mu+\mu_{n-1}(1-\mu), \label{recursion}
\end{eqnarray}
or in the form of one step recursion,
\begin{equation}
\mu_n=\mu(1-\mu_{n-1}) \label{recursion1}
\end{equation}

Although the exact solution of the Bethe anzats  requires a long
technical analysis, the average velocity of the dispersive flow
can be obtained with minimal assumptions from simple combinatorial
arguments. The average velocity of the particle flow in the ASAP
is determined by the average number of steps of all particles
involved into an avalanche during the time interval $t$ and can
be written as
\begin{equation}
v=\frac{\langle Y_t \rangle}{Pt}
\end{equation}
The only assumption we use is that the probability for any site
of the infinitely large lattice to
be occupied does not depend on state of the other sites and is
equal $\rho$. Let us consider the avalanche starting at the site
$i$. To calculate the velocity, we introduce the probability  $P_{i,j}(n)$
for $n$  particles to be transferred from the site $j$ to its right
provided the avalanche is started at site $i$.
Defining also the  total probability to transfer exactly $n$ particles
from any site during the whole avalanche
$P(n)=\sum_{j=i}^\infty P_{i,j}(n)$ we
can express the average velocity as follows
\begin{equation}
v=\sum_{n=1}^\infty nP(n) \label{velos}
\end{equation}
Using the translation invariance of the stationary state one can
rewrite $P(n)$ as the sum of topplings of one site $i$ in all
avalanches started on the left of the site $i$.
\begin{equation}
P(n)=\sum_{j=0}^\infty P_{i-j,i}(n).
\end{equation}
The dynamic rules of the model relate the values of $P(n)$ for two
neighboring sites. Due to the translation invariance, $P(n)$ does not
depend on site, and obeys the recurrent relations
\begin{equation}
 \label{recurrent}
P(n)=P(n-1) \rho \mu_n +P(n)( \rho (1-\mu_{n+1}) +
(1-\rho)\mu_n)+P(n+1)(1-\rho (1-\mu_{n+1})).
\end{equation}
where we express $P(n)$ at the site $i$ via $P(n)$ at the site
$i-1$. For $P(1)$ we have different recurrent relation. The term
with $P(n-1)$ should be replaced by $1$ which corresponds to the
case when avalanche is starting at given site,
\begin{equation}
\label{recurrent1}
 P(1)=1 +P(1)( \rho (1-\mu_{2}))+P(2)(1-\rho) (1-\mu_{2}).
\end{equation}
To find the solution of these recurrent relations one should also
fix the value of $P(1)$. If we want the system to be stationary we
should make sure that the number of starting avalanches is equal
to the number of dying ones. There is $P(1)(1-\rho)$ avalanches
ending at any site for each avalanche starting at the same site,
what gives $P(1)(1-\rho)=1$. Then the solution of the system of
the relations (\ref{recurrent}, \ref{recurrent1}) is
\begin{eqnarray}
P(1)&=&\frac{1}{1-\rho} \\
P(n)&=&\frac{\rho}{1-\rho}\frac{\mu_n}{1-\mu_n}P(n-1),
\end{eqnarray}
or applying these relations recursively
\begin{equation}
 P(n)=\frac{1}{\rho}\left(\frac{\rho}{1-\rho}\right)^n\prod_{i=2}^n
 \frac{\mu_n}{1-\mu_n}.
\end{equation}
Substituting the expression for $P(n)$ into Eq.(\ref{velos}),
and using the recursion (\ref{recursion1}), we get
\begin{equation}
 v=\frac{1}{\rho}\sum_{n=1}^\infty\frac{n}
{\mu_{n+1}}\left(\frac{\mu \rho}{1-\rho}\right)^n
 \label{velosform}
\end{equation}
As we shall see below, this formula  coincides with that we
obtain from the exact Bethe anzats solution. This means that
our assumption about uncorrelated stationary state is valid  in
the thermodynamic limit.

\section{Bethe anzats solution.}
Consider the ASAP consisting of $P$ particles on a ring of $N$
sites and denote by $P_t(C)$ the probability of finding at time
$t$ the system in a configuration $C$. The probability $P_t(C)$
satisfies
\begin{equation}
\label{markovmatrix}
\frac{d}{dt}P_t(C)=\sum_{C^{'}}[M_0(C,C^{'})+M_1(C,C^{'}]
P_t(C^{'})
\end{equation}
where $M_1(C,C^{'})dt$ is the
probability of going from $C^{'}$ to $C$ during the time interval
$dt$, and $M_0$ is a diagonal matrix
\begin{equation}
 M_0(C,C)=-\sum_{C^{'}\neq C}M_1(C^{'},C)
\end{equation}
Before using the Bethe ansatz, it is instructive to note that in
the region where the distances between every two neighbouring
particles exceed $1$, the master equation (\ref{markovmatrix})
becomes ``free'':
\begin{equation}
 \frac{d}{dt}P_t(x_1,...,x_P)=\sum_{k}
e^\gamma P_t(x_1,...,x_k-1,...,x_P)-P_t(x_1,...,x_k,...,x_P)
\label{diffusion}
\end{equation}
where $\exp(\gamma)$ is activity of a single step.
To compensate the difference between (\ref{markovmatrix}) and
(\ref{diffusion}) when $x_k-x_{k-1}=1$ for some $k$, we introduce
the boundary conditions
\begin{eqnarray}
P_t(...,x,x,...)=& e^\gamma(1-\mu)P_t(...,x-1,x,...) + e^{2\gamma}
\mu P_t(...,x-1,x-1,...) \label{boundary}
\end{eqnarray}
This condition can be viewed as the recurrent relation where the
"intermediate" probability of an unstable configuration
$P_t(...,x,x,...)$ is given in terms of another unstable
configuration $P_t(...,x-1,x-1,...)$ and so on. All
boundary conditions for more then two particles can be reduced
to the two-particle case. This implies a recurrent relation for the
probability $\mu_n$ which is nothing but the two particle
reducibility  (\ref{recursion1})  discussed above.

Now, we can define the ASAP by (\ref{diffusion}) and
(\ref{boundary}) instead of (\ref{markovmatrix}) without even
knowing the exact form of the matrix $M(C,C^{'})$ which is very
cumbersome for the ASAP model. Specifying a configuration $C$ by
positions $1\leq x_1<x_2...<x_P\leq N$ of the $P$ particles, we
use the Bethe ansatz for an eigenvector of the matrix $M_0+M_1$ in
the form
\begin{equation}
e^{\lambda t}\sum_{Q} A_{Q}\prod_{j=1}^{P}z_{Q(j)}^{-x_j}
\label{anzats}
\end{equation}
where the sum is over all of the permutations $Q$ of $1,2,...,P$.
The condition (\ref{boundary}) fixes the two particle S-matrix
$A_{ij}/A_{ji}$ as
\begin{equation}
\frac{A_{jk}}{A_{kj}}=-\frac{1-(1-\mu)e^{\gamma}z_{j}
-\mu e^{2\gamma}z_{j}z_{k}}{1-(1-\mu)e^{\gamma}z_{k}
-\mu e^{2\gamma}z_{j}z_{k}}
\end{equation}
Imposing the periodic boundary conditions gives the Bethe equations
\begin{equation}
z_{k}^{-N}=(-1)^{N-1}\prod_{j=1}^{P}\frac{1-(1-\mu)e^{\gamma}z_{j}
-\mu e^{2\gamma}z_{j}z_{k}}{1-(1-\mu)e^{\gamma}z_{k} -\mu
e^{2\gamma}z_{j}z_{k}} \label{betheequations}
\end{equation}
The eigenvalue $\lambda(\gamma)$ corresponding to (\ref{anzats}) is
\begin{equation}
\lambda(\gamma)=-P+e^{\gamma} \sum_{i=1}^{P}z_i
\label{eigenvalue}
\end{equation}
The dependence of the eigenvalue on $\gamma$ allows one to use
$\lambda(\gamma)$ as the large deviation function of $Y_t$.
Specifically, the average velocity of the particle in ASAP is
expressed through the derivative of $\lambda(\gamma)$.
\begin{equation}
\left.
v=\frac{1}{P}\frac{d\lambda(\gamma)}{d\gamma}\right|_{\gamma=0}
\end{equation}

\section{The Bethe  equations in thermodynamic limit.}
The Bethe ansatz equations (\ref{betheequations}) together
with (\ref{eigenvalue}) give the exact solution of the problem
for all $N$ and $P$.
The rest of the paper is devoted to evaluation of $v$ in the
thermodynamic limit $N \rightarrow \infty$ for a fixed density of
particles $\rho=P/N$. For a finite $N$, the largest eigenvalue
$\lambda$ corresponds to the solution $\{ z_j \}$ which converges
to $z_j=1, j=1,...,P$ as $\gamma \rightarrow 0$. For small
$\gamma>0$, the distance $|z_j-1|$ grows rapidly with $N$ for all
$j$ and becomes of order of 1 in the limit $N \rightarrow \infty$.
Introducing a variable $\alpha$ by
\begin{equation}
z_j=\frac{1-e^{i\alpha_j}}{1+\mu e^{i\alpha_j}}e^{-\gamma}
\label{varchange}
\end{equation}
and assuming the solutions $\{\alpha_j\}$ are distributed along a smooth
curve in the complex plane $\alpha=(u+ir)$ with endpoints $(-a+ib)$ and
$(a+ib)$, we obtain the Bethe equation in the form
\begin{equation}
p(\alpha)=2 \pi F(\alpha)+\frac{1}{2\pi}\int_{-a+ib}^{a+ib}
\theta(\alpha-\beta)R(\beta)d\beta - i\gamma \label{integraleq1}
\end{equation}
where we defined as usual a function $F(\alpha)$ such that
$dF/d\alpha=-R(\alpha)/2\pi$ and $F(-a+ib)=-F(a+ib)=\rho/2$.
The functions $p(\alpha)$ and $\theta(\alpha)$ are
\begin{equation}
p(\alpha)=-i\ln
\left(\frac{1-e^{i\alpha}}{1+e^{i\alpha-2\nu}}\right)
\end{equation}
and
\begin{equation}
\theta(\alpha)=-i\ln\big(\frac{\cosh(\nu+i\alpha/2)}{\cosh(\nu-i\alpha/2)}\big)
\label{theta}
\end{equation}
where $\nu=-\ln (\mu)/2$. Taking the derivative in
(\ref{integraleq1}), we get the integral equation for $R(\alpha)$
\begin{equation}
-R(u,b)+\frac{1}{2\pi}\int_{-a}^{a}K(u-v)R(v,b)dv=\xi(u,b)
\label{integraleq2}
\end{equation}
with
\begin{equation}
\xi(\alpha)=\frac{\cosh \nu}{\sinh \nu-\sinh(\nu-i\alpha)}
\end{equation}
and
\begin{equation}
K(\alpha)=\frac{\sinh 2\nu}{\cosh 2\nu+\cos \alpha}
\end{equation}

All that is very similar to the equations for the asymmetric
6-vertex model \cite{Nol,BS}(see also \cite{Sas}) with an
essential exception: both terms containing $z_j$ and $z_iz_j$ in
(\ref{betheequations}) are negative, which is the reason for a
dynamical transition, as we shall show below.

If $a=\pi$, equation (\ref{integraleq2}) can be solved by the
Fourier transformation. To evaluate $\partial_{\gamma}\lambda$, we
have to find the solution of (\ref{integraleq2}) in a vicinity of
the point $a=\pi$ which corresponds to a ``conical'' point,
considered in \cite{BS}. Following Bukman and Shore, we write the
solution $R(u)$ as an expansion in $\epsilon =\pi-a$ up to order
of $O(\epsilon^4)$
\begin{equation}
R(u)=R_0(u)+\epsilon^1\delta R_1(u)+\epsilon^2\delta R_2(u)+
\epsilon^3\delta R_3(u)
\epsilon^4\delta R_4(u)
\label{expansion}
\end{equation}
The  necessity of such a long expansion
will be seen in further calculations.
The Fourier transformation is defined by
\begin{equation}
X(u)=\sum_{n=-\infty}^{\infty}(X)_ne^{-inu}
\end{equation}
where $X$  stands for $R_0,\delta R_m,\xi,K$.
The non-zero Fourier coefficients of $K$ and $\xi$ for $b\geq-2\nu$  are
\begin{equation}
(K)_n=(-1)^{n}e^{-2\nu|n|}
\label{K_n}
\end{equation}
\begin{equation}
(\xi)_n=-e^{bn}(1-(-1)^{n}e^{2n\nu}), n<0
\end{equation}
Then, (\ref{integraleq2}) gives
\begin{equation}
R_0(u)=(R_0)_0+\frac{e^{iu-b}}{1-e^{iu-b}} \label{R_0}
\end{equation}
In the next order in $\epsilon$, we have
\begin{equation}
(\delta R_1)_n(K_n-1)=\frac{R_0(\pi)}{\pi}(-1)^{n}(K)_n
\end{equation}
so, that $R_0(\pi)=0$,$(\delta R_1)_n=0$ and $(R_0)_0=1/(1+\exp
b)$. The next terms in (\ref{expansion}) are evaluated in
\cite{BS}
\begin{equation}
\delta R_2(u)=-\frac{1}{6}R_0^{''}(\pi) \label{R_2}
\end{equation}
\begin{equation}
\delta R_3(u)=-\sum_{n \neq
0}\frac{in(K)_nR_0^{'}(\pi)}{3\pi(1-(K)_n)} e^{-inu} \label{R_3}
\end{equation}
\begin{equation}
\delta R_4=-\frac{1}{120}R_0^{(4)}(\pi)
\label{R_4}
\end{equation}

Thus, in  the expansion of $R(u)$, $\delta R_2(u) \equiv \delta
R_2$ and $\delta R_4(u) \equiv \delta R_4$ are real constants and
$\delta R_3(u)$ is imaginary.

Now, we are ready to start a direct evaluation of
$\partial_{\gamma}\lambda$ in (\ref{eigenvalue}) using
$\partial_{\gamma}\lambda=\partial_{a}\lambda /
\partial_{a}\gamma$. First, we find $\partial_a \gamma$. To this
end, we put $\alpha=a+ib$ in (\ref{integraleq1}), and take the
derivative by $a$ at $a=\pi$. Recalling the conditions
$F(a+ib)=-\rho/2$ and $\theta(0)=0$, we obtain
\begin{eqnarray}
\pi\partial_a \rho+i\partial_a \gamma=& R(a,b)+ (2\pi)^{-1}\theta
(2a)R(-a,b)+\cr &(2\pi)^{-1}\int_{-a}^{a}\theta(a-v)\partial_a
R(v,b)dv \label{normalization}
\end{eqnarray}
To evaluate r.h.s. of (\ref{normalization}), we express the values
$R(a,b)$,$R(-a,b)$ and $\theta(2a)$ by their Tailor expansions at
$a=\pi$ up to the order of $O(\epsilon^3)$ using (\ref{theta}) and
(\ref{expansion}). The integral in (\ref{normalization}) is
treated as
\begin{equation}
\int_{-\pi+\epsilon}^{\pi-\epsilon}f(v)dv=
\int_{-\pi}^{\pi}f(v)dv+B(\epsilon)
\end{equation}
with
$$
B(\epsilon)=\sum_{m=1}^{\infty}\frac{1}{M!}\{(-\epsilon)^m f^{(m-1)}(\pi)-
(\epsilon)^m f^{(m-1)}(-\pi)\}
$$
and, therefore, can be evaluated by the Fourier transformation.
The Fourier coefficients of $\theta (\pi-v)$ are
\begin{equation}
(\theta(\pi-v))_n=\frac{2\pi i}{n}(e^{-2\nu n}-(-1)^n), n\neq 0
\label{theta_n}
\end{equation}
and $(\theta(\pi-v))_0=2\pi^2$. The only $v$-dependent part of
$\partial_a R(v,b)$ is $\delta R_3(v)$. Using
(\ref{K_n},\ref{R_2}-\ref{R_4}), and (\ref{theta_n}), we obtain
the explicit expression for r.h.s. of (\ref{normalization})
\begin{equation}
\epsilon\frac{\pi}{3}R_0^{''}(\pi)
-\epsilon^2\frac{R_0^{'}(\pi)}{\pi}+\epsilon^3\frac{\pi}{30}R_0^{(4)}(\pi)
\label{r.h.s.}
\end{equation}
Due to (\ref{R_0}), the first and third terms in r.h.s. of
(\ref{r.h.s.}) are real, the second one is imaginary. Therefore,
we have
\begin{equation}
\partial_a \gamma=i\epsilon^2\frac{R_0^{'}(\pi)}{\pi}
\label{d_agamma}
\end{equation}

The expression for $\partial_a \lambda$ can be found in a similar way.
In this case, we take the derivative  by $a$  in
\begin{equation}
\frac{\lambda}{N}=\frac{1}{2\pi}\int_{-a}^{a}R(u,b)(z(u,b)-1)du
\end{equation}
where $z(u,b)$, according to (\ref{varchange}), is
\begin{equation}
z(u,b)=\frac{1-e^{iu-b}}{1+e^{iu-2\nu-b}}
\end{equation}
The obtained derivative is similar to (\ref{normalization}) but
contains $z(v,b)$ instead of $\theta(a-v)$. So, we need the
Fourier coefficients of $z(v,b)$ which are
\begin{equation}
(z)_n=(1+e^{2\nu})(-1)^n e^{(2\nu+b)n}, n<0 \label{z_n}
\end{equation}
and $(z)_0=1$.Continuing as in (\ref{normalization}), we get
\begin{equation}
\frac{1}{N}\frac{\partial \lambda}{\partial a}=\epsilon^2
\frac{R_0^{'}(\pi)z^{'}(\pi,b)}{\pi}-3\epsilon^2\sum_{n=1}^{\infty}
(z)_{-n}(\delta R_3)_n \label{d_alambda}
\end{equation}
We can see that the term $\delta R_3(u)$ in (\ref{expansion}) is
relevant. As to $\delta R_4(u)$, it is sufficiently that it is a
constant and does not lead to a divergency by integration.

Substituting the explicit expressions for $(z)_{-n}$ and $(\delta
R_3)_n$ gives the second term in r.h.s. of (\ref{d_alambda}) in
the form
\begin{equation}
\epsilon^2(1+e^{2\nu})\frac{iR_0^{'}(\pi)}{\pi}
\sum_{n=1}^{\infty}\frac{(-1)^n e^{-(4\nu+b)n}}{1-(-1)^n e^{-2n\nu}}
\end{equation}
Due to (\ref{d_agamma}), $R_0^{'}(\pi)$ is cancelling in
$\partial_a \lambda
/
\partial_a \gamma$. Then, using (\ref{z_n}) and
the identity $\rho=(R_0)_0=1/(1+\exp b)$ we obtain the final
result
\begin{equation}
 v=\frac{(1-\rho)(1+\mu)}{(1-\rho(1+\mu))^2}+\frac{1+\mu}{\mu
\rho}\sum_{n=1}^{\infty}\frac{(-1)^n n \mu^{2n}}{1-(-1)^n\mu^n}
\left(\frac{\rho}{1-\rho}\right)^{n},
\label{final}
\end{equation}
which exactly coincides with the formula (\ref{velosform}) after
some algebra.

The velocity of flow $v$ diverges at $\rho_c=1/(1+\mu)$ which
implies a transition to the phase of continuous flow. The value of
critical density $\rho_c$ can be easily understood from the
condition of a balance between gaining $(\rho_c)$ and losing
$(1-q_\infty)$ one particle at each step of a large avalanche.

The considered model is a directed version of the model of activated
random walks introduced in \cite{Vesp} to see how a conservative
dynamical system with the sandpile toppling rules approaches criticality.
It has been shown in \cite{Vesp1} that the relaxation time $\tau$ and
correlation length $\xi$ diverge as $\tau \sim |\rho_c-\rho|^{-\nu_1}$
and $\xi \sim |\rho_c-\rho|^{-\nu_2}$.
The exponents $\nu_1$ and $\nu_2$ have been determined numerically for
several  kinds of toppling rules. In the directed case, $\xi$
coinsides with $\tau$ and is proportional to the average size of avalanches
$\langle s \rangle$. On the other hand, $\langle s \rangle =v$
so we have from (\ref{final}) $\langle s \rangle \sim
(\rho_c-\rho)^{-2}$ and $\nu_1=\nu_2=2$.
An extension of this result to the symmetrical
case \cite{Manna} is a very interesting and difficult problem.
\section{Acknowledgments.}
This work is supported by the grant of Russian Foundation for
Basic Research number 99-01-00882 and by the grant of Swiss
National Science Foundation.  One of us (VBP) is grateful toD.Dhar
for helpful comments. VBP acknowledges T.Dorlas for reading the
manuscript and DIAS, Dublin for hospitality.


\end{document}